\documentstyle[iopl10,epsf]{ioplppt}
\begin{document}
\title{Scaling of level-statistics and critical exponent of disordered 
two-dimensional symplectic systems}
\author{L. Schweitzer$^{\dag}$ and I. Kh. Zharekeshev$^{\ddag}$}
\address{$\dag$ Physikalisch-Technische Bundesanstalt, Bundesallee 100, 
D-38116 Braunschweig, Germany\\
$\ddag$ Institut f\"ur Theoretische Physik, Universit\"at Hamburg, 
Jungiusstra{\ss}e 9, D-20355 Hamburg, Germany}
\begin{abstract}
The statistics of the energy eigenvalues at the metal-insulator-transition 
of a two-dimensional disordered system with spin-orbit interaction is 
investigated numerically. The critical exponent $\nu$ is obtained from the 
finite-size scaling of the number $J_0$ which is related to the 
probability $Q_{n}(s)$ of having $n$ energy levels within an interval of width 
$s$. In contrast to previous estimates, we find $\nu=2.32\pm 0.14$ close to 
the value of the two-dimensional quantum Hall system.   
\end{abstract}

\section{Introduction}
Despite considerable efforts the understanding of the critical behaviour at 
the disorder driven metal-insulator-transition (MIT, Anderson transition) in 
disordered systems is still incomplete.
Present day analytical theories are not able to provide quantitative
results of physical quantities which describe the 
universal properties in the vicinity of the critical point. 
Therefore, our current knowledge mainly originates from 
numerical investigations using transfer-matrix approaches and
Greens-function techniques (see e.g. \cite{KM93}). 
Recently, the energy level statistics has proven to be another 
powerful method to elucidate the peculiar spectral correlations 
and to calculate the non-trivial exponents that govern the 
non-conventional dynamics \cite{CD88,HS94,CKL96,BHS96} that are related 
to the multifractal properties of the corresponding eigenstates. 

In particular, the energy level spacing distribution $P(s)$ is a simple tool
to distinguish between localized and extended states, and also reflects the
respective symmetry of the model system under consideration. 
Here,  $s=|E-E'|/\Delta$ is the energy separation of two consecutive 
eigenvalues $E$ and $E'$ divided by the mean level spacing $\Delta$. 
On the metallic side of the MIT, random-matrix theory (RMT)
serves as an adequate description. However, in approaching the transition, 
novel critical level statistics have been found in three-dimensional (3d) 
\cite{Sea93,ZK95a,BM95,BSZK96,KOSO96} and two-dimensional (2d) 
\cite{SZ95,Eva95,OO95,BS97} systems. 
For small level separations, $s$, these critical $P_c(s)$ still resemble 
the RMT-result, $P(s)\sim s^\beta$, indicating strong level repulsion, 
while for large spacings a behaviour, $P(s)\sim \exp(-\kappa s)$, similar 
to the uncorrelated Poisson-statistics of the localized states is observed. 
Previously, we have found the size-invariant $P_c(s)$ for 2d symplectic 
\cite{SZ95} and QHE systems \cite{BS97} with $\kappa\approx 4.0$, and for
3d orthogonal and unitary models \cite{BSZK96} where $\kappa\approx 1.9$
was obtained.

A disorder driven metal-insulator transition is observed in 3d for each
possible symmetry class: orthogonal ($\beta=1$), unitary ($\beta=2$) and 
symplectic ($\beta=4$). In non-interacting 2d systems, however, in the 
presence of both 
time-reversal and spin-rotational symmetry (orthogonal) all states are 
believed to be localized for any disorder $W>0$. The same holds also in the
absence of time-reversal symmetry (unitary), but singular energies are 
present in strong magnetic fields where the localization length diverges
leading to the quantum Hall effect.

The only complete MIT in 2d is found for symplectic systems that possess 
time-reversal but no spin-rotational symmetry.
Therefore, the symplectic model is of particular interest for developing
analytical theories and the knowledge of characteristic physical quantities 
is of importance. 
Yet, there is no consensus about the magnitude of the universal critical 
exponent $\nu$ that governs e.g. the divergence of the localization length, 
$\xi(E) \sim |E-E_0|^{-\nu}$, at the MIT, whereas the value $\nu\approx 2.35$ 
for the QHE case is commonly accepted \cite{Huc92,Huc95,KHKP91}.
The proposed values for the symplectic system have been found by using 
the transfer matrix method. They are scattered over a wide range: 
$\nu=2.05\pm0.08$ at $W_c/V=5.875\pm 0.010$ \cite{And88,And89}, 
$\nu=2.75\pm 0.15$ at $W_c/V=5.74\pm 0.03$ \cite{Fea91,Fas92}, 
and $\nu=2.5\pm 0.3$ for a different model \cite{Eva95}.
All suggested $\nu$-values are clearly different from the QHE-case which is
contrary to the observed closeness of the generalised fractal dimensions 
$D(q)$ in symplectic and QHE-systems \cite{Sch95}.
Since several recent investigations (e.g. \cite{Mar94,Sch95,SZ95,KO96})
at the MIT were performed taking the numbers ($\nu=2.75$ and critical 
disorder $W_c/V=5.74$) published by Fastenrath \cite{Fea91,Fas92} for granted, 
an independent check of the critical exponent and disorder is necessary.

\section{Model}
Three different models \cite{Mac85,EZ87,And88} have been proposed for  
numerical studies of the localization properties in 2d disordered systems 
with symplectic symmetry.
In the present investigation we use the model suggested by Ando
in order to be able to compare with the above mentioned 
results for the critical exponent. Also, this model seems to be more realistic,
because it simulates transfer of electrons between s-orbitals via p-orbitals
in the presence of spin-orbit interaction.
The latter is responsible for the symplectic symmetry due to the broken
spin-rotational invariance.
In second quantization, the Hamilton operator on a square lattice with sites 
$m$ and $n$ and lattice constant $a$ is
\begin{equation}
H = \sum_{m,\sigma } \varepsilon^{} _{m} c^{\dagger}_{m,\sigma }
c^{}_{m,\sigma} + \sum_{<m\ne n>, \sigma \sigma '} 
V(m,\sigma ; n,\sigma ')\,c^{\dagger}_{m,\sigma } c^{}_{n,\sigma '},
\end{equation}
where the disorder potentials $\varepsilon_{m}$ are random numbers 
distributed between $-W/2$ and $W/2$ with probability $P(\varepsilon)=V/W$.
Periodic boundary conditions are applied in both directions.
The spin-orbit interaction strength $S$ is defined as the ratio 
$S=V_2/(V_1^2+V_2^2)^{1/2}$, where $V_1$ and $V_2$ are matrix elements of 
the $2\times 2$ complex transition matrices $V(m,\sigma ; n,\sigma ')$, 
which depend on the transfer-direction and on spin $\sigma$. 
In the following we choose $S=0.5$ and the unit of energy 
$V\equiv(V_1^2+V_2^2)^{1/2}=1$, and only nearest-neighbour transfer is 
considered. The two-fold degenerate eigenstates are calculated by direct 
diagonalization using a Lanczos algorithm.
\begin{figure}
\epsfxsize12cm\epsfbox{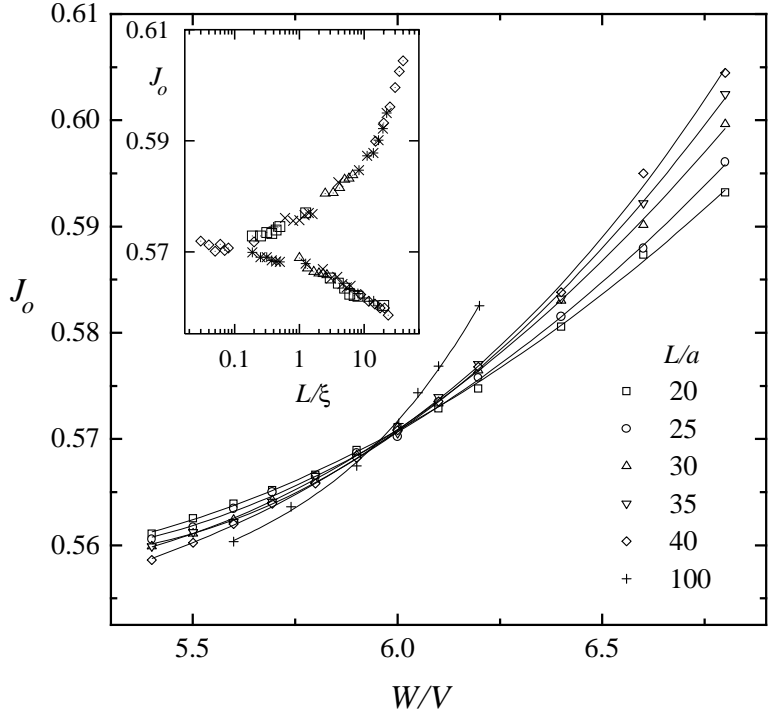}
\caption[]{\label{fig1}The scaling variable $J_0$ versus 
the disorder W for various linear size $L$ of the 2d square lattice.
The inset: the two-branch scaling curve $J_0(L/\xi)$ showing the
metal-insulator transition at $J_0^c \approx 0.571$.
Different symbols correspond to different disorder.}  
\end{figure}
\section{Results and Discussion}
The eigenvalue statistics are calculated in a given interval 
[$-0.5\,V$, $0.5\,V$] around the band centre $E/V=0.0$.
A careful spectral unfolding procedure was applied to compensate for possible
global variations in the density of states which would disturb the interesting
local correlations. More than $10^6$ eigenvalues
were accumulated for each of the parameter pairs $(L,W)$ with linear system 
size $L/a=15$, 20, 25, 30, 35, 40, 100 and disorder strength ranging
from $W/V=5.4$ to 6.8 by calculating up to 3000 realizations of the 
disorder potentials.
As a scaling variable, we consider the quantity~\cite{ZK95b}
\begin{equation}
J_0(L,W) = \int_{0}^{\infty} Q_{0}(s) ds, 
\end{equation} 
where $Q_{n}(s)$ for $n=0$ is the probability that an energy interval 
of width $s$ contains no energy eigenvalue. It is related to the 
nearest-neighbour level spacing distribution, 
$P(s) \equiv d^2 Q_{0}(s)/ds^2$~\cite{Meh91}. 
Compared to previous approaches
\cite{Sea93,HS94,ZK95a} the major advantage of our choice is that all 
calculated neighbouring level spacings  are used, because no arbitrary 
cut-off parameter is needed.  

The spectral correlations of a disordered system undergo
a crossover from the Wigner to the Poisson statistics when increasing
the disorder~\cite{Sea93,BM95,ZK95a}.
In the metallic phase, $J_0 \approx 0.522$ is well known for infinite 
symplectic systems from RMT \cite{Meh91}, and in the insulating 
phase from the Poisson distribution $J_0=1$.
As expected, the level statistic $J_0$ as a function of $W$ changes 
from the RMT-result ($W<W_c$) to the Poisson limit ($W>W_c$)
continuously for finite $L$, but discontinuously in the thermodynamic limit.
It exhibits critical behaviour close to the disorder $W_c$, which separates
the extended and localized regimes. The sign of the finite-size 
effect is reversed when crossing the fixed point, signalizing the 
delocalization-localization transition.

Results of $J_0(L,W)$ as a function of disorder are plotted in Fig.\,1 for six
different system sizes. A point of intersection at $W/V \approx 5.98$ is
clearly observed. Within numerical error $\delta W \approx 0.04$, 
it is consistent with the value reported in~\cite{And89}, but certainly 
larger than that from~\cite{Fea91}.  
We find our $J_0(L,W)$-results to fall in between the 
limiting values mentioned above with an size independent 
$J_0^c = 0.571\pm0.001$ at the critical point.
Similar behaviour is also found for $J_{n \ge 1}(L,W)$ by calculating 
the probability $Q_n(s)$ of the non-zero number of eigenvalues $n$ in 
the interval $s$.
We obtained a scale-invariant sequence of the critical numbers $J_n^c$,
which grow with $n$, $J_1^c \approx 0.976$, $J_2^c \approx 0.996$, ...,
and converge very fast to unity. In 3d orthogonal systems, an analogous 
critical sequence \cite{ZK95b} was found earlier. 

Assuming the validity of the one-parameter scaling hypothesis, 
$J_0(L,W)=f(L/\xi(W))$, one can introduce the correlation length $\xi(W)$
and re-scale the linear size $L$ so that  the data of $J_0$ as a function 
of $L/\xi$ for consecutive $W$ overlap with each other. In approaching the 
critical point the precision of $\xi$ decreases due to less overlap.
Within statistical accuracy all points fall onto one common curve 
independently of $L$ and $W$. This yields the two-branched scaling function 
displayed in the inset of Fig.\,1, where $J_0(L,W)$ is plotted versus 
$L/\xi(W)$.
The lower and upper branches indicate the extended and localized sides of a 
complete metal-insulator transition, respectively.

\begin{figure}
\epsfxsize7.5cm\epsfbox{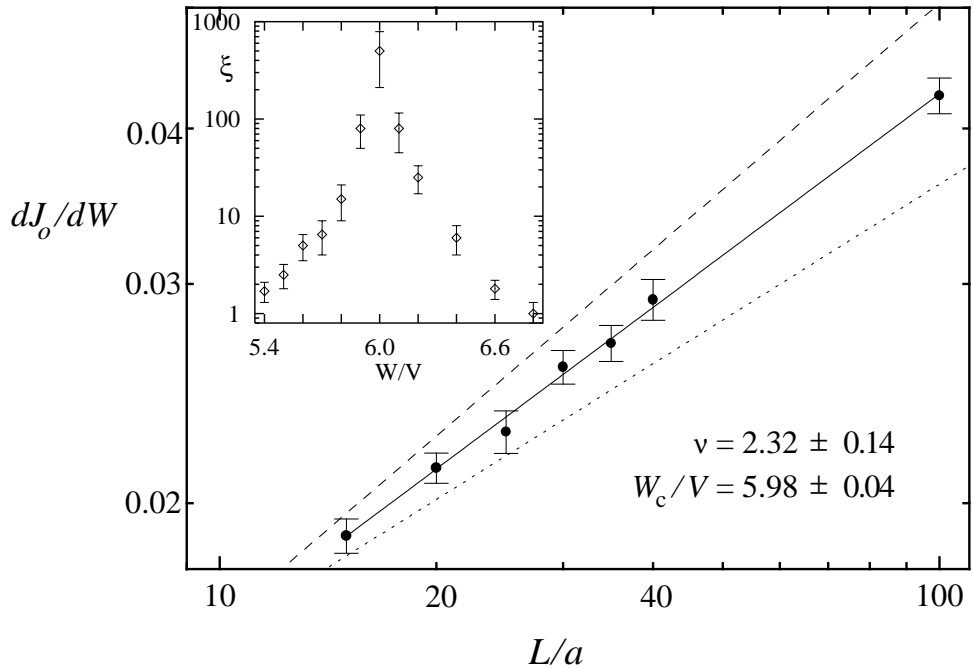}
\caption[]{\label{fig2}
The derivative $dJ_0/dW$ at $W_c$ as a function of the system size $L$ 
(log-log plot). The solid line is a linear fit to the raw data. Dashed 
and dotted lines correspond
to the values of the critical exponent $\nu=2.05$ and $2.75$ obtained
in~\cite{And89} and~\cite{Fea91}, respectively.
The inset shows the disorder dependence of the correlation length $\xi(W)$.}
\end{figure}

In order to calculate the critical exponent $\nu$ one can expand
the finite-size scaling function near $W=W_c$, 
\begin{equation}
J_0(L,W) \approx J_0^c + A (L/\xi(W))^{1/\nu}=J_0^c + 
A^{\prime} (W-W_c)L^{1/\nu}.
\label{expansion}
\end{equation}
This gives  $dJ_0(L,W)/dW \propto L^{1/\nu}$. As soon as $W_c$ is detected
one can perform the two-parameter linear fit on double-log scale, the inverse
slope being $\nu$. Fig.\,2 shows the derivative $dJ_0(L,W)/dW$ at $W_c=5.98$
as a function of the system size $L$ from which a critical exponent 
$\nu=2.32\pm0.14$ can be extracted. Recent calculations of a symplectic 
network model yield a similar exponent \cite{Jan}.
As a guide to the eye, two additional lines are drawn for comparison,
the slopes of which correspond to the values of $\nu$ obtained by Ando 
\cite{And88,And89} and Fastenrath \cite{Fea91,Fas92}, using the 
transfer-matrix method.  

If $W_c$ is not known with sufficient accuracy, a four-parameter fitting 
procedure to Eq.~(\ref{expansion}) is required. 
By applying a $\chi^2$-criterion similar to that from~\cite{Huc91,Mac94}, 
we verified that the resulting $W_{c}$ and $\nu$ do not change considerably.
However, their uncertainties increase as expected, $\delta W = 0.10$ and 
$\delta \nu = 0.16$. Thus, our value of $\nu$ is clearly distinct from those 
obtained previously. Note that the obtained error-bar $\delta \nu$ does not 
overlap with corresponding error-bars of those $\nu$ from previous studies.
The inset of Fig.\,2 displays the correlation length $\xi$ as a function of 
disorder $W$. Here, $\xi$ is defined up to a constant factor $\xi_0$.
It diverges at the critical disorder in agreement with the 
power law $\xi(W)=\xi_0 |W-W_c|^{-\nu}$. 
It is generally believed that the value of $\nu$ is universal and does not 
depend on the strength $S$ of the spin-orbit interaction. It remains to be 
checked whether this expectation really holds. 

In conclusion, we have investigated the critical properties of the energy 
level statistics at the metal-insulator transition of a disordered 
two-dimensional system with symplectic symmetry. The scaling function and 
the critical exponent were calculated using the statistics of spacings
of neighbouring energy eigenvalues that have been obtained numerically. 
By performing a finite-size scaling analysis we found a critical exponent 
$\nu=2.32$ at a critical disorder $W_c/V=5.98$ which is markedly different 
from those values reported previously. Our value of $\nu$ for the symplectic 
system is very close to that found in 2d QHE-systems, a behaviour that has 
been observed previously also for the generalized fractal dimensions.  
\section*{References}
\bibliographystyle{prsty}

\end{document}